\documentclass[prl,reprint,superscriptaddress]{revtex4-1}
\usepackage{graphicx}
\usepackage{xcolor}
\usepackage{amsmath}
\usepackage{amssymb}
\usepackage[caption=false]{subfig}
\usepackage{gensymb}
\usepackage{appendix}
\usepackage{hyperref}

\begin{document}

\author{Arabinda Bera}
\email{arabinda.bera@unimi.it}
\affiliation{Department of Physics ``A. Pontremoli", University of Milan, via Celoria 16, 20133 Milan, Italy}

\author{Debjyoti Majumdar}
\altaffiliation{Present address: RIKEN Center for Biosystems Dynamics Research, Kobe, Japan}

\affiliation{Alexandre Yersin Department of Solar Energy and Environmental Physics, Jacob Blaustein Institutes for Desert Research,\\
Ben-Gurion University of the Negev, Sede Boqer Campus 84990, Israel}

\author{Ido Regev}
\affiliation{Alexandre Yersin Department of Solar Energy and Environmental Physics, Jacob Blaustein Institutes for Desert Research,\\
Ben-Gurion University of the Negev, Sede Boqer Campus 84990, Israel}

\author{Timothy W. Sirk}
\affiliation{Polymers Branch, US Army DEVCOM Army Research Laboratory, Aberdeen Proving Ground, Maryland 21005, USA}

\author{Alessio Zaccone}
\thanks{\color{blue}alessio.zaccone@unimi.it\color{black}}
\affiliation{Department of Physics ``A. Pontremoli", University of Milan, via Celoria 16, 20133 Milan, Italy}

\title{Microscopic origin of shear bands in 2D amorphous  solids from topological defects}
\begin{abstract}
The formation of shear bands in amorphous solids such as glasses has remained an open question in our understanding of condensed matter and amorphous materials. Unlike in crystals, well-defined topological defects such as dislocations have been elusive due to the lack of a periodic ordered background at the atomic level. Recently, topological defects have been identified in the displacement field and in the eigenvectors of amorphous solids. Recent work has suggested that shear bands in amorphous solids coincide with an alignment of vortex-antivortex dipoles, with alternating topological charge +1/-1. Here we numerically confirm this hypothesis by means of well-controlled simulations in 2D. Surprisingly, we show that a chain of topological defects (TDs) pre-exists the shear band and is visible already in the non-affine displacement field of the elastic regime. This chain is activated into a flow band concomitantly with the disappearance and possibly annihilation of a dipole at a distance from the TDs chain. The possible underlying mechanism is reminiscent of a soliton-like rarefaction pulse remotely activated by dipole annihilation as observed in superfluid Bose-Einstein condensates.
\end{abstract}

\maketitle

Understanding the rheology of amorphous solids has been one of the main goals of the fields of soft matter physics as well as of engineering disciplines \cite{falk2011deformation,nicolas2018deformation,cubuk,nesterenko}. One of the main challenges in this field is understanding the yielding transition from elastic to plastic deformation \cite{balmforth2014yielding,bonn2017yield,xu2021microscopic}. In recent years, it was shown that the yielding point exhibits several hallmarks of a nonequilibrium phase transition \cite{parisi2017shear,jaiswal2016mechanical,dahmen2011micromechanical}. It is also known that, depending on the material preparation protocol, yielding occurs either by large, diffuse avalanches, or narrow, localized shear-bands \cite{balmforth2014yielding,ozawa2018random,schall2010shear,fielding2014shear}. In crystalline materials, shear-bands are persistent structures in the sense that once they are formed, they do not spontaneously vanish. In these materials, plastic deformation is mainly mediated by dislocations -- topological line-defects in the crystalline structure -- which are easily recognizable from the material structure. This is not the case in amorphous solids where shear-bands are a result of intermittent slip-patterns that occur on the same band \cite{shi2006strain}. Contrary to the well-defined structural signature of strain localization in crystals \cite{wang2018strain}, a shear-ban long does not have an easily identifiable structural signature in amorphous solids, though indirect measures such as long-range strain correlations \cite{Chikkadi}, machine-learned structural descriptors \cite{PhysRevMaterials.6.065602}, and the potential energy or ``disorder temperature'', were shown to be correlated with the location of the shear band \cite{falk2011deformation,shi2007atomic,manning2009strain}. It is known that shear-bands in amorphous solids are composed of a collection of localized plastic events, referred to as ``soft spots'' \cite{manning2011vibrational} or ``shear transformation zones'' \cite{argon1979plastic,falk1998dynamics,bouchbinder2009nonequilibrium}, which cause an elastic displacement that can be described using the Eshelby inclusion construction \cite{dasgupta2012atomic,Wilde} and its ramifications \cite{moshe}. Recent work has suggested that the displacement field of the shear band can also be described as arising from topological defects \cite{rosner2024} and it was shown that some of these defects are correlated with the positions of soft spots \cite{desmarchelier2024plasticity, bera2025burgers}. Here we show that topological defects (TDs) exist not only in the displacement fields of plastic events but also in the non-affine displacement fields of the “elastic branches” (regions of linear-elastic response in the stress-strain curve) and that these topological defects induce a shear-band-like structure in the elastic displacement field.

Topology underlies the mathematical description of a wide range of natural phenomena by providing a framework that connects microscopic interactions to emergent macroscopic behaviors through invariant global properties. Among others, one key marker is given by TDs, which arise due to incommensurate ordering between the local structure and the underlying lattice symmetry. While it is straightforward to conceptualize TDs in ordered solids like crystals, their definition in disordered systems remained elusive for a long time \cite{sharon}. Only recently, Baggioli et al. \cite{PhysRevLett.127.015501} for the first time discovered the existence of TDs in disordered solids as well-defined toplogically-invariant singularities in the non-affine part of the microscopic displacement field. Two years later, Wu et al. \cite{wu2023topology} first observed discretized TDs, vortices and anti-vortices, in the eigenvectors of the Hessian matrix of 2D glasses. Since then TDs have been observed in various simulated \cite{bera2024,desmarchelier2024plasticity} and also experimental systems \cite{Vaibhav2025}. The discretised TDs have been formulated also for 3D glasses in terms of hedgehog TDs which come in two flavours, radial and hyperbolic \cite{bera2025}. The flavour is also playing an important role in deciding the correlation between the TDs and the soft spots for plasticity \cite{bera2025}. However, the connection between topological defects and shear band formation, i.e. one of the main mechanisms of plasticity in amorphous materials, remains restricted to theoretical predictions only. For instance, in Ref. \cite{rosner2024} the sequential alignment of TDs of alternating charges along the shear band was hypothesized, which is able to explain the characteristic asymmetric sinusoidal oscillation of density along the shear band direction observed experimentally \cite{PhysRevB.95.134111}. However, no such physical evidence has yet been observed in numerical simulations. To validate this hypothesis, we numerically investigate the presence of topological defects in model glass formers, specifically, in ultrastable glasses where a clear, distinct shear band is realizable and typically observed \cite{ozawa2018}. We observe, indeed, that shear bands form in correspondence of spatial alignments of TDs of alternating charge, as first suggested in \cite{rosner2024}. More surprisingly, the chain of aligned TDs is always observed already in the elastic deformation regime, i.e. well before the shear band is formed (see Supplementary Material, SM4). The triggering of the flow band appears to be closely connected with the sudden disappearance of a dipole of TDs. A possible mechanism for shear band formation due to TDs dynamics is presented.

To prepare well-annealed ultrastable glasses we use the swap Monte Carlo algorithm on a polydisperse collection of discs following a power law diameter $(\sigma)$ distribution $P(\sigma)\sim \sigma^{-3}$. The particles interact via a purely repulsive potential. For details see Supplementary Material SM1 and SM2. For most of our analysis we use samples consisting of $N=4096$ particles, prepared at $T=0.1$ unless specified otherwise. To check any finite size effects we also reproduce the results in $N=1024$ particles system. Once equilibrated, we instantaneously quench the system to its inherent state using the Fast Inertial Relaxation Engine (FIRE) algorithm, which is then subjected to uniform athermal quasistatic shear (see SM3). While shearing, each incremental deformation $\delta \gamma=10^{-4}$ is followed by an additional step of structural relaxation using FIRE.


We first look at the stress-strain curve under uniform shear and verify that the ultrastable glasses yield differently than glasses prepared at higher temperatures (such as $T=1$); this is substantiated by the sharp stress drop for the former, which is absent for the latter. The yielding point is located to be at $\gamma_c = 0.0787$. 
We checked that our observations remain consistent across 30 independent samples. 

\begin{figure}[h]
\centering
\includegraphics[width=\linewidth]{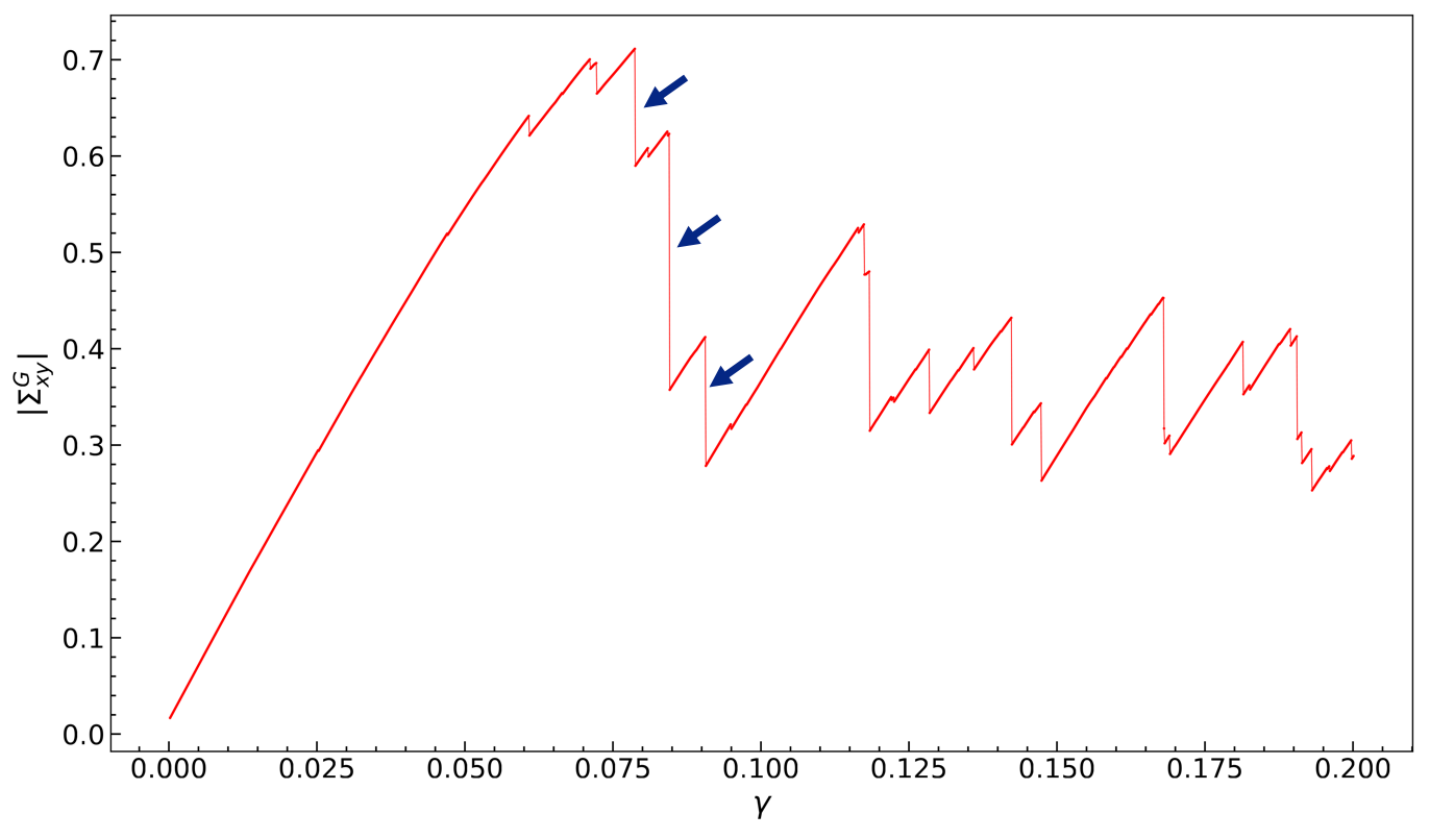}
\caption{The stress-strain curve for a representative 2D glass is shown. Three major stress drops are indicated by the arrows.}
\label{fig1}
\end{figure}

\begin{figure*}[ht]
\centering
\includegraphics[width=\linewidth]{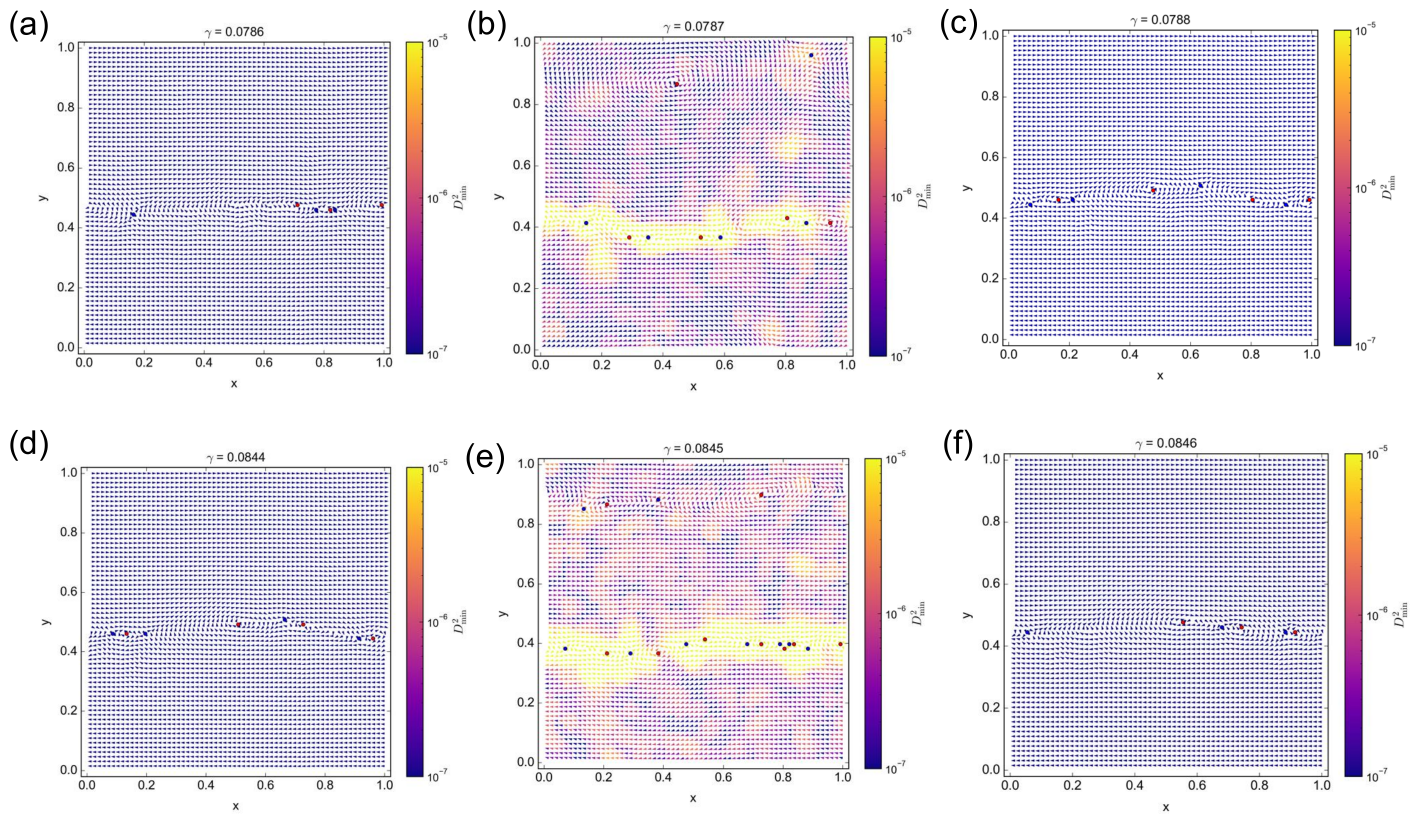}
\caption{Normalized displacement fields around two plastic events, at $\gamma=0.0787$ and $0.0845$, as indicated on the stress-strain curve in Fig. \ref{fig1}, are shown using vector arrows. The topological point defects in the displacement field are identified: vortices with topological charge $q = +1$ (red) and anti-vortices with $q = -1$ (blue). The vectors are color-coded by the local non-affine displacement measure $D^2_{\min}$. An extended, system-spanning, shear band is observed in panels (b) and (e) near the middle of the simulation box, indicating localized plastic flow at $\gamma = 0.0787$ and at $\gamma=0.0845$, respectively. However, the aligned chain of TDs pre-exists the flow band (panels (a) and (d)) and a similar chain is still observed after the stress drop (panels (c) and (f)). }
\label{fig2}
\end{figure*}

Figure \ref{fig1} shows a typical stress–strain curve, where we highlight three major stress drops corresponding to plastic events. For each of these stress drops, we analyze the particle displacement field and compute the local non-affine deformation using the $D^2_{\min}$ metric (non-affine displacements are particle motions in addition to those dictated by the macroscopic strain tensor \cite{zaccone2011approximate}). As shown in Fig. \ref{fig2}, a well-developed shear band forms along the shear direction during these events. We also identify topological point defects in the displacement field, with vortices ($q=+1$) and anti-vortices ($q=-1$) localized primarily within the shear band. These defects exhibit a sequential arrangement of opposite charges positioned in the middle of the shear band, as first suggested in the phenomenological model of Ref. \cite{rosner2024}.

A striking feature is the appearance of a large defect dipole far from the shear band during each stress drop. As we could verify by analyzing tens of different simulations, this dipole, which is often isolated, never occurs in the absence of a plastic stress drop, suggesting that it may serve as a signature of long-range elastic interactions triggering the plastic flow in the shear band.

These observations support three key insights:\\
(i) The spatial organization of topological defects in the displacement field along with the isolated defect dipole can serve as a marker for plastic events and reveal the presence of extended shear bands, even before their occurrence.\\
(ii) The role of dipoles far from the shear band remains unclear but it appears to be connected to stress redistribution and to nonlinear energy transmission leading to plastic flow (see below in what follows for further discussion).\\
(iii) The emergence of isolated dipoles accompanying the onset of the shear band is robust with respect to changes in the system size and moderate changes in the interaction potential, indicating the possible universality of their role in triggering the shear band.\\

\begin{figure*}[ht]
\centering
\includegraphics[width=\linewidth]{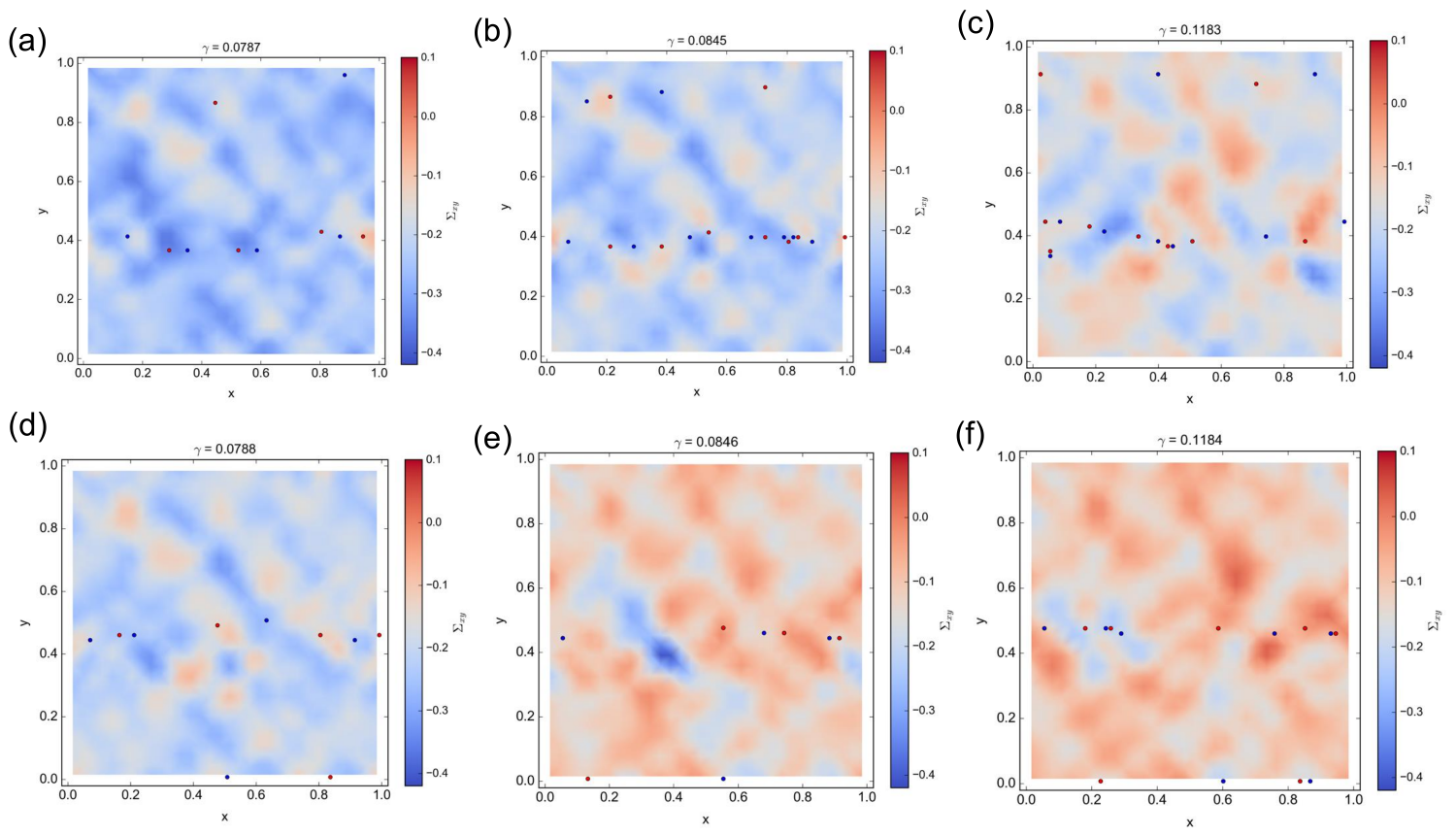}
\caption{Local shear stress field $\Sigma_{xy}$ is shown right before (top row (a)-(c)) and right after (bottom row (d)-(f)) the three plastic events marked in Fig. \ref{fig1}. A pronounced change is observed, with large ribbon-like regions of the sample transitioning from negative to slightly positive stress or viceversa, highlighting the spatial redistribution of internal stresses associated with the plastic rearrangement. The locations of TDs with $q=+1$ (red) and $q=-1$ (blue) are overlaid on the stress field.}
\label{fig3}
\end{figure*}

In Fig. \ref{fig3}, we show the virial stress maps in the system before and after the stress drops (the same as indicated in Fig. \ref{fig1}). As is evident, the compressive negative stress concentrates in 135 degrees-ribbon like patterns, as expected for force chains under shear deformation. After the plastic event (stress drop), the sign of the stress in these ribbon-like structures flips from negative (blue) to positive (red). The sign reversal or ``stress recoil'' effect can be understood as follows, by means of a simple 1D model. Let \( x \in [-L/2, L/2] \) be the coordinate across the material. A shear band of thickness \( h \ll L \) is centered at \( x = 0 \).

The total imposed shear displacement is \( \Delta u \), and the displacement field is \( u(x) \). The total strain as imposed externally, is conserved and can be written as:
\[
\gamma(x) = \frac{du}{dx} = \gamma_e(x) + \gamma_p(x)
\]
where:  $ \gamma_e(x)$ is the elastic shear strain,
   $ \gamma_p(x) $ is the plastic shear strain, localized in the shear band $\gamma_p(x) = 0 $ for \( |x| > h/2 \)). Let us suppose that plastic deformation localizes in the band:
\begin{align}
\gamma_p(x)& \gg 0 \quad \text{for} \quad |x| < \frac{h}{2},\nonumber\\
\gamma_p(x) &= 0 \quad \text{for} \quad |x| > \frac{h}{2}.\nonumber
\end{align}
To maintain compatibility with the externally imposed displacement:
\[
\int_{-L/2}^{L/2} \left[ \gamma_e(x) + \gamma_p(x) \right] dx = \Delta u
\]
that is, the elastic strain in a certain region outside the shear band must change its sign — and hence also the corresponding local stress must undergo a sign reversal. This simple argument explains the ``recoil effect" of the local stress at a plastic event, shown in Fig. \ref{fig3}.

Let us now revert to the shear band formation mechanism, in Fig. \ref{fig2}. As mentioned above, the appearance of the flow band is always accompanied by the presence of a dipole (often isolated) at a distance from the alignment of TDs that acts as precursor of the shear band.
Tens of different simulations have been carefully checked, and this same situation has been observed in all of them, with no exception. Also, it has been impossible to observe the dipole in the absence of a concomitant shear band. We also note that the topological charge separation distance ("dipole moment"), between the +1 and the -1 topological charges in the dipole, is always quite significant (covering a significant fraction of the simulation box size). In turn, this implies that the dipole carries a large energy. Importantly, we always observe that the dipole disappears as soon as the shear band is set in motion. It is very likely that the vortex-antivortex (VA) dipole annihilates right before the shear band begins to flow. While we still cannot prove it directly due to the limited resolution in strain of our analysis, we can at least provide a \emph{reductio ad absurdum} argument. Indeed, if we were to assume that the dipole does not annihilate (in the very narrow strain increment during which the shear band starts to flow), we would still observe it even after the plastic event and the flow band, which is in contradiction with all our evidence collected on tens of different independent simulations.

The striking concurrence of a VA dipole disappearance and possibly, annihilation, and of the flow band is highly reminiscent of a phenomenon which is well known in superfluid Bose-Einstein condensates (BECs) made with cold atoms or photons \cite{PhysRevX.1.021003,PhysRevA.90.063627,Prabhakar_2013,PhysRevLett.134.233401}. In these systems, it is ubiquitously observed that the annihilation of a VA dipole triggers the emission of a solitonic rarefaction pulse (a propagating local dip in density). This phenomenon is well described by the Gross-Pitaevskii equation, which is mathematically identical to the nonlinear Schr\"{o}dinger equation used to describe the nonlinear mechanics of wave propagation in solids \cite{sulem,nesterenko}. Hence, we can speculate that, in glasses, something similar may happen: the annihilation of a topological VA dipole releases a quote of kinetic energy which is funneled into the emission of a rarefaction pulse. The latter is most likely to occur at the pre-existing chain of TDs, because this is the most suited place to act as a nonlinear waveguide. In turn, the rarefaction pulse that propagates along the TDs chain triggers the flow band, because a propagating local dip in density directly causes the local plastic slip and the flow of the particles. 

This mechanism is largely plausible, and can  explain various features of shear banding in amorphous solids, e.g. the finite thickness of the shear band (due to the localization of the rarefaction pulse), the sinusoidal density oscillation along the shear band (explained, again, by the rarefaction pulse), and the bursting, intermittent character of the shear bands.  

In summary, we have numerically demonstrated, by means of well-controlled MD simulations in 2D, that shear banding in glasses originates from the dynamics of topological defects in the microscopic displacement field. The shear bands always occur on pre-existing alignments of TDs with alternating charge +1/-1 (vortex-antivortex). Importantly, these aligned chains of TDs pre-exist the onset of the shear band, and are observed already for infinitesimal elastic deformations (see SM4). This finding opens up the way for the unprecedented possibility to accurately predict the locations of shear bands in amorphous materials without the need to simulate their plastic deformation. The intriguing disappearance of a vortex-antivortex dipole, with a large dipole moment, typically located far from the aligned TDs chain, provides a possible link with the physics of vortex-antivortex annihilation in superfluids.

\section*{Acknowledgments}
We acknowledge discussion and input with Davide E. Galli and Woojin Kwon. DM was supported by the BCSC Fellowship from the Jacob Blaustein Center for Scientific Cooperation. IR and DM were supported by the Israel Science Foundation grant no. 1204/23. AB and AZ gratefully acknowledge funding from the US Army DEVCOM Army Research Office through contract nr.  W911NF-22-2-0256. AZ gratefully acknowledges funding from the European Union through Horizon Europe ERC Grant number: 101043968 “Multimech”.
In the making of this paper, we became aware of similar observations in a 2D experimental granular system \cite{wang_2025}, that appeared a few days before this preprint. 

\section*{Author contribution}
DM and AB contributed equally. DM prepared the glass samples and their deformed configurations under uniform shear. AB performed the topological defects analysis. AZ and IR supervised the work. AZ wrote the paper with input from the other co-authors.

\bibliography{bib.bib} 


\section*{Supplementary material }

\section{SM1: Model \label{appendix_model}}
We model our glass formers as two-dimensional discs interacting via a repulsive potential \cite{berthier2019} of the form: 

\begin{equation}
U_{ij}(r)= \epsilon_0 \left( \frac{\sigma_{ij}}{r}  \right)^{12} + c_0 + c_1 \left(\frac{r}{\sigma_{ij}}  \right)^{2} + c_2 \left( \frac{r}{\sigma_{ij}} \right)^{4}
\end{equation}
for $r\leq r_c=1.25\sigma_{ij}$ and $0$ otherwise, where $\sigma_{ij}$ is the mean-diameter giving the center-to-center distance, and is calculated as
\begin{equation}
\sigma_{ij} = \frac{\left( \sigma_i + \sigma_j \right)}{2}(1-\epsilon \left( \sigma_i - \sigma_j \right)).
\label{eq_diameter}
\end{equation}
The constant $\epsilon_0$ sets the units for energy and temperature  (with $k_B=1$) and is set to $\epsilon_0=1$. In Eq. \ref{eq_diameter}, $\epsilon$ controls the degree of non-additivity, thereby enhancing grouping of differently sized particles promoting the glass forming ability and suppressing crystallization and is set to a value of $0.2$. The constants $c_0$, $c_1$ and $c_2$ are chosen such that the potential, its first- and second-derivatives vanish at $r_{cut}/\sigma_{ij}=1.25$. 


Additionally, to enhance the efficacy of the swap-MC algorithm, we chose a continuous size polydispersity where the particle diameters $(\sigma)$ follow a power law distribution, $P(\sigma)\sim \sigma^{-3}$, with $\sigma \in \left[ \sigma_{\text{min}}, \sigma_{\text{max}} \right]$ and chosen as $\sigma_{\text{min}}=0.72$ and $\sigma_{\text{max}}=1.6$. The degree of size polydispersity can be measured from $\delta \sigma = \sqrt{\bar{\sigma^2}-\bar{\sigma}^2}/\bar{\sigma}$ and is found to be $\delta \sigma \approx 0.23$ which follows from the fact that $\sigma_{\text{min}}/\sigma_{\text{max}}=0.45$. The average diameter $\bar{\sigma}$ sets the unit of length. We simulate a system of number density $\rho=N/V=1$ where $N$ is the number of particle and $V$ is the volume (area of the simulation box in 2D).

\section{SM2: Sample preparation \label{appendix_sample}}
To prepare ultra-stable glasses we anneal our glass formers at low temperatures using the swap-Monte Carlo (SMC) method along with initial molecular dynamics (MD) run. To start with we prepare samples at a relatively high temperature $T=3.0$, and successively equilibrate it at two other lower temperatures at $T=2$ and $T=1$, each for $50$ MD time units. The final MD equilibrated samples are then used as the starting configuration for SMC. Thereafter, we follow the steps below to equilibrate the system using SMC: 
\begin{itemize}
    \item[(a)] a particle $i\in \{1,2,...,N\}$ is randomly chosen from the set of all particles $(N)$.  
    \item[(b)] we decide if swap is to be performed with probability `$p_s$', otherwise we try only translating the chosen particle in a random direction. We chose a slightly increased value of $p_s=0.2$ in comparison to previous works \cite{grigera2001}. 
    \item[(c)] in case swap is to be performed, we choose another random particle $\{j\in\{1,2,...,N\}\mid j\neq i\}$ from the set of particles excluding the first one, and check if the difference in diameters $(\delta \sigma = \mid \sigma_i - \sigma_j \mid)$ between the two chosen particles $\delta\sigma  \leq 0.2$. This extra condition ensures better acceptance rate since particles chosen from two extremities of the distribution are geometrically incommensurate and is unlikely to be accepted under the Metropolis criterion. If $\delta\sigma  \leq 0.2$ is satisfied, we swap the radii of the chosen particles.
    \item[(d)] following step (b) or (c), the chosen particle/s is/are given a random displacement of magnitude $\delta r=10^{-3}$.  
    \item[(e)] finally, the new configuration is accepted only if the energy change satisfies $\mathcal{R} \leq \exp(-\beta \triangle E)$, where $\beta=1/T$, $\triangle E = E_{\text{new}}-E_{\text{old}}$ and $\mathcal{R}$ is a uniform random number $\mathcal{R} \in [0,1]$.
\end{itemize}

    The value of $\delta r$ is chosen such that swap acceptance probability remains substantial. Too large values of $\delta r$ can lead to most of the trials being rejected, while too small a value makes the equilibration process lengthy. We consider a sample to be sufficiently equilibrated if the mean energy do not changes over the last $10^5$ MC steps, where each MC step comprises $N$ trials. Finally, the SMC equilibrated samples are quenched to its inherent state using the FIRE algorithm, which is then used to perform uniform shear and defect analysis.

\section{SM3: Athermal quasistatic shearing}
Athermal quasistatic shearing along the +x direction is performed by applying Lee-Edwards boundary condition along the y-direction, while the $x=0$ and $x=L$ boundaries have regular periodic boundary condition. Further, each step of infinitesimal deformation is followed by an energy minimization using the FIRE algorithm. For most purposes we have chosen an incremental deformation of $\delta \gamma = 10^{-4}$ with the exception of $\delta \gamma = 10^{-7}$ near the yielding point when specified.

\section{SM4: Topological defects analysis \label{appendix_topological_defects}}
We compute the displacement field of each particle $i$ as $u_i(\gamma)=\vec{r}_i(\gamma+\delta \gamma)-\vec{r}_i(\gamma)$, where $\vec{r}_i$ denotes the position of particle $i$. This displacement field is interpolated onto a $64 \times 64$ regular grid. We then identify topological defects by computing the topological charge $q$ associated with the phase angle $\theta$ of the vector field, using the discretized form of the line integral:
\begin{equation}
    q=\frac{1}{2 \pi} \oint_{\mathcal{L}} d\theta,
\end{equation}
where $\mathcal{L}$ is a closed loop encircling a grid cell. We evaluate this winding number by considering the phase differences at the four corners of each square grid. The defect core is assigned to the center of the square grid. The topological charge $q$ typically takes integer values, such as $q=-1,0,+1$ corresponding respectively to anti-vortices, neutral regions, and vortices.

\begin{figure*}[h]
\centering
\includegraphics[width=\linewidth]{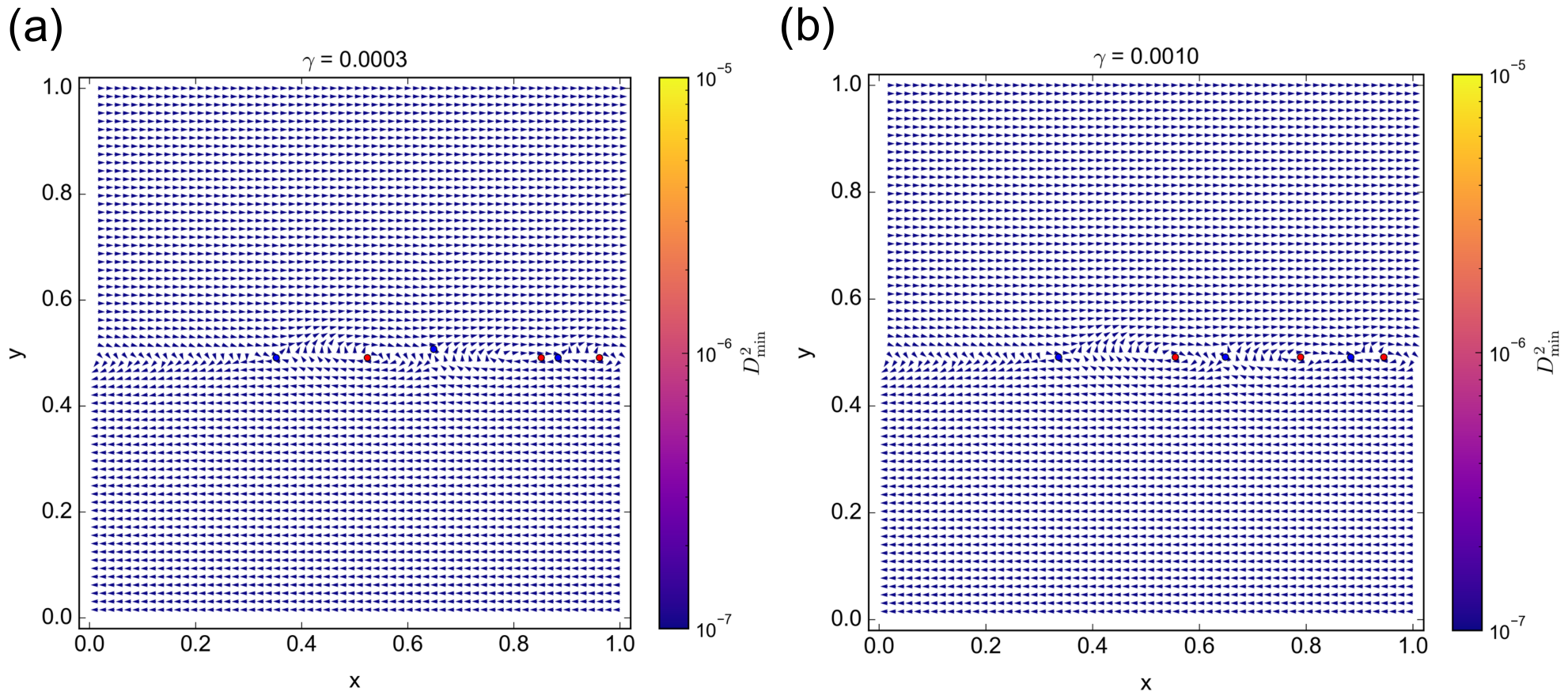}
\caption{(a)-(b) Displacement field patterns in the elastic regime at strains $\gamma=0.0003$ and $0.001$. The topological defects (TDs) exhibit alternating charge alignment and maintain similar spatial configurations across strain values in elastic region.}
\label{sm_fig1}
\end{figure*}

We identify the topological defects within the displacement field of the glassy sample for the entire range of $\gamma$ by considering a small strain step $\delta \gamma =10^{-4}$. Interestingly, even within the nominally elastic regime, we observe spatially correlated arrangements of defects with alternating topological charge. These structures appear robust and repeatable across different strain values, as illustrated in Fig.~\ref{sm_fig1}.

\section{SM5: Computation of Global and Local Stress \label{appendix_stress_calc}}
During the athermal quasistatic (AQS) simulations, we compute the instantaneous global virial stress tensor, volume-averaged over the entire simulation box. The global stress tensor is defined as
\begin{equation}
    \Sigma^{G}_{\alpha \beta} = -\frac{1}{2V}\sum_{i,j}(r^{\alpha}_i-r^{\alpha}_j)f^{\beta}_{ij};
\end{equation}
where $r^{\alpha}_i$ is the $\alpha$-component of the position vector of particle $i$, and $f^{\beta}_{ij}$ is the $\beta$-component of the interparticle force exerted on particle $i$ by particle $j$. The summation runs over all interacting pairs $(i, j)$ in the system, and $V$ denotes the total volume of the simulation domain. The global stress-strain response is characterized by plotting $|\Sigma^G_{xy}|$ as a function of the applied shear strain $\gamma$ as shown in Fig. \ref{fig1}.

In order to capture the spatial variations in stress, we further compute a coarse-grained local stress tensor $\Sigma_{\alpha \beta}(\vec{r}_c)$ on each grid point. We begin by evaluating the pairwise stress contributions:
\begin{equation}
\tau^{ij}_{\alpha \beta} = - (r_i^\alpha - r_j^\alpha) f_{ij}^\beta.
\end{equation}
Each contribution $\tau^{ij}_{\alpha \beta}$ is assigned to the midpoint $\vec{r}_{ij,m} = (\vec{r}_i + \vec{r}_j)/2$ of the interacting pair.

We partition the simulation box into a $32 \times 32$ uniform grid and compute the local stress at each grid cell centered at $\vec{r}_c$ by considering a Gaussian weighting function, defined as
\begin{equation}
W(\vec{r}_c - \vec{r}_{ij,m}) = \exp\left( -\frac{|\vec{r}_c - \vec{r}_{ij,m}|^2}{2\sigma^2} \right),
\end{equation}
where $\sigma$ is the coarse-graining width. The local stress tensor at position $\vec{r}_c$ is then computed as
\begin{equation}
\Sigma_{\alpha\beta}(\vec{r}_c) = \frac{\sum_{i>j} W(\vec{r}_c - \vec{r}_{ij,m}) \tau^{ij}_{\alpha\beta}}{\sum_{i>j} W(\vec{r}_c - \vec{r}_{ij,m})},
\end{equation}
with the summation performed over all interacting pairs $(i, j)$ that contribute non-zero stress.



\end{document}